\documentclass[prl,aps,showpacs,twocolumn,superscriptaddress,nofootinbib]{revtex4-1}
\usepackage{amsmath, amsthm, amssymb}
\usepackage{subfigure}
\usepackage{graphicx}
\usepackage{array}
\usepackage{multirow}
\usepackage[colorlinks=true,linkcolor=blue,citecolor=blue,urlcolor=blue]{hyperref}
\usepackage[separate-uncertainty,retain-explicit-plus,per-mode=symbol,range-phrase=\mbox{--},range-units=single]{siunitx}
\usepackage{rotating}
\usepackage{placeins}
\usepackage{upgreek,xspace}
\usepackage{lineno}

\usepackage{stackengine}
\usepackage[percent]{overpic}
\usepackage[symbol*]{footmisc}
\DefineFNsymbolsTM{otherfnsymbols}{%
  \textdagger    \dagger
  \textasteriskcentered *
  \textsection   \dagger
  \textbardbl    \|%
  \textparagraph \mathparagraph
  \textbullet \circ
  \textdaggerdbl \ddagger
}%
\setfnsymbol{otherfnsymbols}

\usepackage{wasysym}
\usepackage[separate-uncertainty,retain-explicit-plus,per-mode=symbol,binary-units]{siunitx}
\usepackage{array,mathtools,amssymb,dcolumn}
\usepackage{tikz}
\usepackage{multirow}
\usepackage{afterpage}
\usepackage{lineno}
\usepackage{paralist}
\usepackage{listings}
\usepackage{array}
\usepackage{cancel}
\usepackage{xspace}
\UseRawInputEncoding
\usepackage{stmaryrd}
\usepackage[version=4]{mhchem}

\newcommand{\Geant}{\texttt{Geant4}}

\DeclareSIUnit\c{\mbox{$c$}}
\DeclareSIUnit\week{w}
\DeclareSIUnit\year{yr}
\DeclareSIUnit\yr{yr}
\DeclareSIUnit\day{day}
\DeclareSIUnit\standard{std}
\DeclareSIUnit\str{sr}
\DeclareSIUnit\ppm{ppm}
\DeclareSIUnit\ppb{ppb}
\DeclareSIUnit\ppt{ppt}
\DeclareSIUnit\pe{PE}
\DeclareSIUnit\spe{SPE}
\DeclareSIUnit\ev{events}
\DeclareSIUnit\hit{hit}
\DeclareSIUnit\hits{hits}
\DeclareSIUnit\bin{(\mbox{5-PE}~bin)}
\DeclareSIUnit\sgm{\mbox{$\sigma$}}
\DeclareSIUnit\rms{RMS}
\DeclareSIUnit\keVr{\mbox{keV$_{\rm nr}$}}
\DeclareSIUnit\keVee{\mbox{keV$_{\rm ee}$}}
\DeclareSIUnit\MeVee{\mbox{MeV$_{\rm ee}$}}
\DeclareSIUnit\ph{photons}
\DeclareSIUnit\pm{PMT}
\DeclareSIUnit\inch{''}
\DeclareSIUnit\bit{bit}

\DeclareSIUnit\sample{S}
\DeclareSIUnit\barn{b}
\DeclareSIUnit\bara{bar}
\DeclareSIUnit\Curie{Ci}
\DeclareSIUnit{\msun}{\mbox{M$_\odot$}}
\DeclareSIUnit\mK{\milli\kelvin}
\DeclareSIUnit\micron{\micro\metre}
\DeclareSIUnit\liveday{\mbox{live-days}}
\DeclareSIUnit\tonneday{\mbox{tonne$\cdot$days}}
\DeclareSIUnit\tonneyear{\mbox{tonne$\cdot$years}}
\DeclareSIUnit\days{\mbox{days}}

\usepackage[symbol*]{footmisc}
\DefineFNsymbolsTM{otherfnsymbols}{%
  \textdagger    \dagger
  \textasteriskcentered *
  \textbardbl    \|%
  \textparagraph \mathparagraph
  \textbullet \circ
  \textdaggerdbl \ddagger
}%
\setfnsymbol{otherfnsymbols}

\newcommand{\NeutrinoPaperTwoEndDate}{March 27, 2020}
\newcommand{\NeutrinoPaperLiveTime}{815}
\newcommand{\NeutrinoPaperExposure}{\SI{7.29\,\pm\,0.05}{\tonneyear}}

\newcommand{\nuObs}{5}
\newcommand{\nuObsExtended}{6}

\newcommand{\nuSigExtended}{4.0\,$\sigma$}
\newcommand{\nuSigBGExtented}{1.5\,$\sigma$}  
\newcommand{\nuSigExpectedboroneight}{$2.3\,\pm\,0.4$}
\newcommand{\nuSigExpectedhep}{$0.03\,\pm\,0.01$}

\newcommand{\effectiveXCsixNobsROI}{$(4.0~^{+2.0}_{-1.6}~\mathrm{(stat)}~^{+0.8}_{-0.7}~\mathrm{(sys)})\times 10^{-41}\,\mathrm{cm}^2$}
\newcommand{\effectiveXCBhatt}{\SI[tight-spacing=true]{1.67\pm0.05e-41}{\square\cm}}
\newcommand{\effectiveXCscale}{$(2.4~^{+1.3}_{-1.0})$}

\newcommand{\aboveOneDelayedT}{\SI{252}{\,\ns}}

\newcommand{\radiogenicBG}{$0.43\,\pm\, 0.05{\mathrm{\,(stat)}}\,\pm\, 0.14{\mathrm{~(sys)}}$}
\newcommand{\radiogenicBGMC}{$0.22$}
\newcommand{\radiogenicBGBAT}{$0.15$}

\newcommand{\cosmogenicTotal}{$0.048\,\pm\,0.009{\mathrm{~(stat)}} {}^{~+0.070} _{~-0.025}~{\mathrm{~(sys)}}$}
\newcommand{\corrGamGam}{$<10^{-3}$}

\newcommand{\corrAlphaAlpha}{$<10^{-3}$}
\newcommand{\uncorrAlphaAlpha}{$<10^{-3}$}
\newcommand{\totBackground}{$0.48~^{+0.16}_{-0.15}$}

\newcommand{\LArLateTime}{\SI{1.6}{\micro\second}}
\newcommand{\nratePMT}{\SI{14.7}{\milli\becquerel}}

\newcommand{\nuROI}{\SIrange{10.5}{13.0}{\MeV}}
\newcommand{\nuROIEnu}{\SIrange{12.0}{14.5}{\MeV}}

\newcommand{\muonFluxSNO}{$(3.31\pm0.01_{stat}\pm0.09_{sys})\times10^{-10} \mathrm{\mu/cm^{2}/s}$}

\newcommand{\neutrinoFluxSNOthreePhase}{$(5.25\,\pm\,0.16\,{\mathrm{(stat)\,}}^{+0.11}_{-0.13}\,{\mathrm{(sys)}})\times 10^{6}\mathrm{/cm^{2}/s}$}
\newcommand{\neutrinoFluxSNOthreePhaseSym}{\SI[tight-spacing=true,per-mode=reciprocal]{5.2\pm0.2e6}{\per\square\cm\per\second}}
\newcommand{\neutrinoFluxhep}{\SI[tight-spacing=true,per-mode=reciprocal]{0.8\pm0.2e4}{\per\square\cm\per\second}}

\newcommand{\branchingMetastable}{$29\,\%$}

\newcommand{\larmasserror}{\SI{3269\,\pm\,24}{\kg}} 
            
\newcommand{\larlevel}{\SI{551}{\mm}}

\newcommand{\PaperTwoStartDate}{November 4, 2016}

\newcommand{\kforty}{\mbox{$^{40}$}K}

\newcommand{\arforty}{\mbox{$^{40}$Ar}}

\newcommand{\beight}{\mbox{$^{8}$B}}
\newcommand{\hep}{\mbox{\textit{hep}}}

\newcommand{\DEAP}{\mbox{DEAP-3600}}

\newcommand{\MV}{\mbox{MV}}

\newcommand{\DSk}{\mbox{DarkSide-20k}}
\newcommand{\Argo}{\mbox{ARGO}}

\newcommand{\PMTs}{\mbox{PMTs}}

\newcommand{\alpds}{\mbox{$\alpha$-decays}}

\newcommand{\gr}{\mbox{$\gamma$-ray}}
\newcommand{\grs}{\mbox{$\gamma$-rays}}
\newcommand{\alphan}{\mbox{($\alpha,n$)}}
\newcommand{\ngamma}{\mbox{($n,\gamma$)}}

\newcommand{\PE}{\mbox{PE}}

\begin{document}
\title{First evidence of neutrino absorption on argon using \beight\ solar neutrinos in DEAP-3600}
\newcommand{\Alberta}{Department of Physics, University of Alberta, Edmonton, Alberta, T6G 2R3, Canada}
\newcommand{\AstroCeNT}{AstroCeNT, Nicolaus Copernicus Astronomical Center, Polish Academy of Sciences, Rektorska 4, 00-614 Warsaw, Poland}
\newcommand{\BHSU}{School of Natural Sciences, Black Hills State University, Spearfish, SD 57799, USA}
\newcommand{\Cagliari}{Physics Department, Universit\`a degli Studi di Cagliari, Cagliari 09042, Italy}
\newcommand{\UCR}{Center for Experimental Cosmology \& Instrumentation, Department of Physics and Astronomy, University of California, Riverside, CA 92521, USA}
\newcommand{\CNL}{Canadian Nuclear Laboratories, Chalk River, Ontario, K0J 1J0, Canada}
\newcommand{\Carleton}{Department of Physics, Carleton University, Ottawa, Ontario, K1S 5B6, Canada}
\newcommand{\CIEMAT}{Centro de Investigaciones Energ\'eticas, Medioambientales y Tecnol\'ogicas, Madrid 28040, Spain}
\newcommand{\Houston}{Department of Physics, University of Houston, Houston, TX 77204, USA}
\newcommand{\INFNCagliari}{INFN Cagliari, Cagliari 09042, Italy}
\newcommand{\INFNNapoli}{INFN Napoli, Napoli 80126, Italy}
\newcommand{\Mainz}{Institut f\"ur Kernphysik, Johannes Gutenberg-Universit\"at Mainz, 55128 Mainz, Germany}
\newcommand{\LNGS}{INFN Laboratori Nazionali del Gran Sasso, Assergi (AQ) 67100, Italy}
\newcommand{\LU}{School of Natural Sciences, Laurentian University, Sudbury, Ontario, P3E 2C6, Canada}
\newcommand{\LBNL}{Nuclear Science Division, Lawrence Berkeley National Laboratory, Berkeley, CA 94720, USA}
\newcommand{\UNAM}{Instituto de F\'isica, Universidad Nacional Aut\'onoma de M\'exico, A.\,P.~20-364, Ciudad de M\'exico~01000, Mexico}
\newcommand{\Napoli}{Physics Department, Universit\`a degli Studi ``Federico II'' di Napoli, Napoli 80126, Italy}
\newcommand{\Capodimonte}{Astronomical Observatory of Capodimonte, Salita Moiariello 16, I-80131 Napoli, Italy}
\newcommand{\MEPhI}{National Research Nuclear University MEPhI, Moscow 115409, Russia}
\newcommand{\Oxford}{Department of Physics, University of Oxford, Oxford, OX1 3PU, United Kingdom}
\newcommand{\IIPA}{International Institute for Particle Astrophysics, Polish Academy of Sciences,  Bartycka 18, 00-716 Warsaw, Poland}
\newcommand{\Princeton}{Physics Department, Princeton University, Princeton, NJ 08544, USA}
\newcommand{\Queens}{Department of Physics, Engineering Physics, and Astronomy, Queen's University, Kingston, Ontario, K7L 3N6, Canada}
\newcommand{\RHUL}{Royal Holloway University London, Egham Hill, Egham, Surrey TW20 0EX, United Kingdom}
\newcommand{\RAL}{Rutherford Appleton Laboratory, Harwell Oxford, Didcot OX11 0QX, United Kingdom}
\newcommand{\SL}{SNOLAB, Lively, Ontario, P3Y 1M3, Canada}
\newcommand{\Sussex}{University of Sussex, Sussex House, Brighton, East Sussex BN1 9RH, United Kingdom}
\newcommand{\TRIUMF}{TRIUMF, Vancouver, British Columbia, V6T 2A3, Canada}
\newcommand{\TUD}{Institut f\"ur Kern und Teilchenphysik, Technische Universit\"at Dresden, 01069 Dresden, Germany}
\newcommand{\TUM}{Department of Physics, Technische Universit\"at M\"unchen, 80333 Munich, Germany}
\newcommand{\MI}{Arthur B. McDonald Canadian Astroparticle Physics Research Institute, Queen's University, Kingston, ON, K7L 3N6, Canada}

\affiliation{\Alberta}
\affiliation{\AstroCeNT}
\affiliation{\BHSU}
\affiliation{\Cagliari}
\affiliation{\UCR}
\affiliation{\CNL}
\affiliation{\Carleton}
\affiliation{\CIEMAT}
\affiliation{\Houston}
\affiliation{\INFNCagliari}
\affiliation{\INFNNapoli}
\affiliation{\Mainz}
\affiliation{\LU}
\affiliation{\UNAM}
\affiliation{\Capodimonte}
\affiliation{\Napoli}
\affiliation{\MEPhI}
\affiliation{\Oxford}
\affiliation{\IIPA}
\affiliation{\Queens}
\affiliation{\RHUL}
\affiliation{\RAL}
\affiliation{\SL}
\affiliation{\TRIUMF}
\affiliation{\TUD}
\affiliation{\TUM}
\affiliation{\MI}

\author{P.~Adhikari}\affiliation{\Carleton}
\author{P.-A.~Amaudruz}\affiliation{\TRIUMF}
\author{D.\,J.~Auty}\affiliation{\Alberta}
\author{M.~Batygov}\affiliation{\LU}
\author{B.~Beltran}\affiliation{\Alberta}
\author{M.\,A.~Bigentini}\affiliation{\Carleton}\affiliation{\MI}
\author{C.\,E.~Bina}\affiliation{\Alberta}\affiliation{\MI}
\author{W.~Bonivento}\affiliation{\INFNCagliari}
\author{M.\,G.~Boulay}\affiliation{\Carleton}
\author{J.~Brachman}\affiliation{\Queens}\affiliation{\MI}
% ...

\collaboration{DEAP Collaboration}
\altaffiliation{deap-papers@snolab.ca}

\date{\today}

\begin{abstract}
We report experimental evidence for electron neutrino charged-current interactions (neutrino absorption, CC~$\nu_e$) from \beight\ solar neutrinos on \arforty\ using an exposure of \NeutrinoPaperExposure\ in the DEAP-3600 detector. A region of interest (ROI) of \nuROI\ reconstructed energy calibrated on single-peak events, corresponding to incident neutrino energy in \nuROIEnu, is used for this measurement. We observe \nuObs\ single-peak and 1 double-peak neutrino-like events consistent with the \beight\ solar neutrino energy spectrum in the ROI after correcting for nonlinearities in the detector response at high energies. With an expected background of \totBackground\ events, the data correspond to a significance of \nuSigExtended\ with respect to the background-only hypothesis. We report an energy-averaged cross section of \effectiveXCsixNobsROI\ in the ROI for the CC~$\nu_{e}$ signal, a factor \effectiveXCscale\ higher than predicted by Bhattacharya, Goodman and Garc\'ia (2009).
\end{abstract}
\maketitle

R.~S.~Raghavan proposed that electron neutrino absorption on \arforty\ ($\nu_e + \arforty \rightarrow \mathrm{\kforty}^* + e^-$) could be used as a new approach to observe low-energy neutrino interactions with matter~\cite{PhysRevD.34.2088}. Raghavan's model considered only the super-allowed 0\textsuperscript{+}$\rightarrow$0\textsuperscript{+} transition from the ground state of \ce{^40Ar} to the \SI{4.38}{\MeV} excited state of \kforty, which de-excites via \gr\ emission. Recent experiments on the \ce{(p,n)} reaction on \arforty\ have revealed additional Gamow-Teller (GT) transitions to other states~\cite{PhysRevC.80.055501}. This is particularly important, as including these additional allowed transitions lowers the reaction threshold from \SI{5.9}{\MeV} to \SI{3.8}{\MeV} and boosts the interaction cross section approximately three-fold. In the low-energy limit where unbound nuclear final states can be neglected, the kinematics of the reaction are well constrained and the neutrino energy is given by 
\begin{equation}
    E_{\nu} = E_{e} + E_{\gamma} + Q = E_{\textrm{vis}} +  \SI{1.5}{\MeV},
\label{eq:raghavan_kinematics}
\end{equation}
where $E_{\nu}$ is the incident neutrino energy, $E_{e}$ is the final-state electron energy, $E_{\gamma}$ is the sum of all de-excitation \grs\ emitted from the \kforty\ nucleus, $E_{\textrm{vis}}$ is the total energy measured directly from the prompt electron and de-excitation \grs\ and $Q=\SI{1.5}{\MeV}$ is the Q-value. The de-excitation cascade passes through a \SI{1.64}{\MeV} excited state with a branching ratio of \branchingMetastable, which is a metastable state with a  lifetime of \SI{480}{\nano\second} producing a delayed coincidence signature~\cite{GARDINER2021108123}.

In this letter, we report the first evidence for electron neutrino charged-current absorption on \arforty\ with \beight\ solar neutrinos, using the DEAP-3600 detector. 
This is also
the first observation of a charged-current neutrino interaction in a dark matter detector. As next-generation multi-tonne scale detectors come online, this signal could be measured with higher precision enabling an improved measurement of the \beight~neutrino spectral shape, direct detection of solar \hep\ neutrinos~\cite{capozzi2019dune} and measurement of the neutrino spectrum from supernovae~\cite{abi_supernova_2021}. 
Additionally, detectors with sensitivity to the energy and multiplicity of final-state \grs\ ({\it e.g.} DarkSide-20k~\cite{Aalsethetal2018}, \Argo~\cite{DS20K_ARGO}, DUNE~\cite{capozzi2019dune}) may be able to further constrain the transition strengths and more accurately compute the interaction cross section.

  Located at SNOLAB, 
  \DEAP~\cite{deap3600_231day} is a dark matter direct detection experiment using \larmasserror\ of liquid argon (LAr)~\cite{Adhikari:2023ces}. 
   A detailed description of the detector is provided in Ref.~\cite{deap3600_design}.
 The scintillation light produced by particle interactions in argon is detected with 255 Hamamatsu R5912-HQE photomultiplier tubes (PMTs)~\cite{AMAUDRUZ2019373} attached to light guides. 
 A \SI{4.8}{\kilo\becquerel} americium-beryllium (AmBe) neutron source is deployed during calibration runs to obtain the energy response from (n,$\gamma$) reactions on detector materials.

The charge response of the PMTs is individually calibrated as described in Ref.~\cite{AMAUDRUZ2019373}. In a given event, the sum of PMT charges observed, expressed in units of the mean single photoelectron charge, corresponds to a number of photoelectrons ($\textrm{PE}$). We correct for light yield changes over time on a run-by-run ($\sim$daily) basis using the measured energy of the 1.46~MeV \gr\ arising from the radioactive decay of \kforty. 
For $E_{\textrm{vis}} > \SI{2.5}{MeV}$, the number of $\textrm{PE}$ and the energy resolution $\sigma$ are related to $E_{\textrm{vis}}$ using a quadratic model~\cite{Rethmeier:2021oez}:
\begin{gather}
\begin{aligned}
 \textrm{PE}(E_{\textrm{vis}})&=\textrm{C} + \textrm{B}\cdot E_{\textrm{vis}} +  \textrm{A}\cdot E_{\textrm{vis}}^{2}, \\
 \sigma^2&=(1+\sigma^2_{\text{PE}})\cdot\textrm{PE}+\sigma^2_{\text{rel, LY}}\cdot\textrm{PE}^2,
\end{aligned}
\label{eq:erespModel}
\end{gather}
where $\textrm{C}=\SI{549}{\pe}$, $\textrm{B}=\SI{6854}{\pe\per\MeV}$ is the light yield of the detector, $\textrm{A}=\SI{-19.51}{\pe\per\MeV^2}$ is a quadratic correction term that accounts for the non-linear response of the detector at higher energies,
$\sigma^2_{\text{PE}}=\SI{2.97}{\pe}$ is a resolution scaling factor that accounts for effects such as the Fano factor and \PE\ counting noise, and ${\sigma^2_{\text{rel, LY}}=3.0\times 10^{-5}}$ accounts for the variance of the light yield relative to its mean value.
For each event, the reconstructed energy is obtained from the number of PE by applying the reciprocal of Eqn.~\ref{eq:erespModel}.

Data used in this analysis were collected between \PaperTwoStartDate\ and \NeutrinoPaperTwoEndDate\ corresponding to 
an exposure of \NeutrinoPaperExposure. 
Run selection similar to the analysis presented in Ref.~\cite{deap3600_791day} is applied to the dataset, with the inclusion of 37 additional runs recovered in the context of this analysis.
Event selection cuts are applied to minimize the backgrounds in the region of interest (ROI). 
The signal acceptance is a measure of the fraction of predicted neutrino events that survive event selection cuts, applying in the denominator a number of basic cuts (\texttt{Low-level cuts}) that select electronic recoil events. 
Muon-induced backgrounds are removed by cutting events that occur within [-0.15,~1.35] ms of an event in the muon veto (\MV). Muon events are identified by requiring at least \SI{10}{\pe} in the \MV\ spread over at least 3~\MV\ \PMTs\, and that the light distribution in the water tank is sufficiently diffuse to be consistent with a muon track: these cuts were designed to accept $\geq$96\% of muons that pass through the shield water tank while suppressing spurious \MV\ triggers due to dark noise~\cite{AndrewThesis}. 
This requirement (\texttt{Muon veto cut}) is taken into account as a loss in live time instead of a loss in signal acceptance.

The DEAP-3600 detector also has four PMTs mounted around the neck coupled to optical fibers designed to tag events that originate in the neck of the detector.
We require the fraction of the total light collected in these neck veto PMTs to be less than \SI{0.5}{\percent} of the total light seen by the LAr PMTs (\texttt{Neck veto cut}).
We also require that the reconstructed vertical position of the event is below the LAr fill level, located \larlevel\ above the equator, to ensure that the events originate within the LAr (\texttt{Reconstructed height cut}).   
Further to this, we select events with a time of the main peak in the waveform that is well-aligned with the hardware trigger, between 2250--2700~ns (\texttt{Event timing cut}), non-noisy events with fewer than four early pulses before the main peak (\texttt{Early pulse suppression cut}), and a maximum of \SI{20}{\percent} of the total charge collected coming from the PMT with maximum charge (\texttt{PMT charge distribution cut}).
In order to remove pile-up events while accepting both prompt and delayed \grs\ from the neutrino signal, we select events with a maximum of two peaks in the waveform (\texttt{Pile-up cut}).
To ensure that all selected events are consistent with LAr scintillation, where the photon time distribution is dominated by the LAr triplet state with a lifetime of \LArLateTime\ microseconds \cite{deap2020pulseshape}, for events with reconstructed energy \mbox{$>\SI{10}{\MeV}$} we fit the slow component of the scintillation waveform and select events with time constants in the range of \SIrange{1.4}{1.8}{\micro\second} (\texttt{Scintillation waveform cut}).
The signal acceptance is studied with high-energy \grs\ from neutron capture events in the neutron calibration data. The cumulative signal acceptance after each selection cut is listed in Table~\ref{tab:cut_acceptance}, along with the number of events remaining in the ROI after each cut is successively applied.

\begin{table}[h]
    \centering
    \caption{Live time loss and expected signal acceptance from event selection cuts (relative to the low-level cuts) displayed in a cumulative way. The number of ROI events passing the series of cuts is shown at each step.
    The same 6 events are cut by the reconstructed height cut as by the neck veto cut.}
    \begin{tabular}{ccc}
    \hline\hline
        \textbf{Event selection cut} &  \textbf{Live time} &  \textbf{ROI Events}\\
        \hline
\texttt{Low-level cuts} & 817.9~d & 12\\
\texttt{Muon veto cut} & 814.7~d & 12\\
\hline
        \textbf{Event selection cut} &  \textbf{Acceptance} &  \textbf{ROI Events}\\

\hline
\texttt{Neck veto cut} & 1.00 & 6\\
\texttt{Reconstructed height cut} & 0.99 & 6\\
\texttt{Event timing cut} & 0.98 & 6 \\
\texttt{Early pulse suppression cut} & 0.97 & 6 \\

\texttt{PMT charge distribution cut} & 0.97 & 6 \\
\texttt{Pile-up cut} & 0.95 &  6 \\
\texttt{Scintillation waveform cut} & 0.92 & 6\\
    \hline\hline
    \end{tabular}
    \label{tab:cut_acceptance}
\end{table}

The range of the ROI in reconstructed energy is \nuROI. The ROI was selected to sufficiently reduce backgrounds while optimizing the sensitivity to solar neutrinos assuming the nominal neutrino flux and cross-section described later. The low end of the ROI is primarily affected by radiogenic background, the upper end by cosmogenic backgrounds. The ROI was defined before looking at the data.
Taken into account in the ROI definition is that the energy response function, calibrated on single-peak events, systematically reconstructs double-peak events at higher energies: for a double-peak event with components $E_1$ and $E_2$, we note $\mathrm{PE}(E_1)+\mathrm{PE}(E_2)>\mathrm{PE}(E_1+E_2)$ due to Eqn.~\ref{eq:erespModel} being non-linear. 
Consequently, the reciprocal of Eqn.~\ref{eq:erespModel} yields a higher reconstructed energy for double-peak events than for single-peak events with the same total deposited energy in the LAr.
Noting that $\mathrm{PE}^{-1}(~\mathrm{PE(11.4~MeV)+PE(1.6~MeV)~)}=13.2~\mathrm{MeV}$, we accept double-peak events up to equivalent single-peak reconstructed energy of 13.2 MeV into the ROI.

\begin{table}
    \centering
    \caption{Backgrounds in the neutrino search ROI.}
    \resizebox{\linewidth}{!}{
    \begin{tabular}{cc}
    \hline\hline
       {\bf Background Source}  & {\bf Events in ROI in \NeutrinoPaperLiveTime~live-days}\\
       \hline
       Radiogenic \ngamma\ & \radiogenicBG \\
       Cosmogenic  & \cosmogenicTotal \\
       Correlated $\gamma-\gamma $ pile-up & \corrGamGam  \\
       Uncorrelated $\gamma-\gamma $ pile-up  & \corrGamGam  \\
       Correlated $\alpha-\alpha $ pile-up  & \corrAlphaAlpha  \\

       Uncorrelated $\alpha-\alpha $ pile-up &  \uncorrAlphaAlpha \\
       \hline
       \textbf{Total Background}  & \totBackground  \\
    \hline\hline
    \end{tabular}
    }
    \label{tab:backgrounds_table}
\end{table}

The backgrounds are listed in Table \ref{tab:backgrounds_table}. The dominant background in the ROI arises from radiative neutron capture on isotopes with Q-values~\mbox{$>\SI{10}{\MeV}$}, creating \ngamma\ events in the ROI. These neutrons are primarily produced by spontaneous fission and \alphan\ reactions in different materials in the detector~\cite{EmmaThesis}. The dominant neutron sources include the PMTs, the stainless steel vessel and the water tank.  
 A total of 60~live-days of neutron calibration data from the AmBe source are used to estimate this background, corresponding to $\sim$\num{e7}\,live-days of background neutrons in the physics dataset assuming a neutron rate of \nratePMT\ from the PMT glass~\cite{PhysRevD.100.072009}. 
 After applying the same selection criteria as listed in Table~\ref{tab:cut_acceptance}, the selected AmBe data are re-scaled to match the integral of the physics dataset in the \SIrange{8.0}{10.0}{\MeV} range. This portion of the spectrum is primarily dominated by \ngamma\ capture events. The background estimate using this method is obtained by integrating the scaled AmBe spectrum in the neutrino ROI. Some effects are unique to or more pronounced in the AmBe data, such as a higher rate of pile-up events, non-thermal captures, and neutron inelastic scatters followed by captures within the same data acquisition window. These are quantified and corrected for in the background estimates.

To evaluate systematic differences between neutrons from the AmBe source and those produced in detector materials during physics data collection, a Monte Carlo (MC) study is performed using the Reactor Analysis Tool (RAT) \cite{RAT}, which interfaces with  \Geant\ (version 4.10.7)~\cite{agostinelli2003geant4} and \texttt{G4RiversideCascade}~\cite{g4cascade}, a package for \Geant\ to more accurately model \ngamma\ de-excitation cascades. Two MC samples were generated to account for different spatial and energy distributions between the calibration and physics data: an AmBe simulated dataset, in which neutrons were generated at the AmBe calibration positions, and a PMT simulated dataset, in which neutrons were generated in the PMTs with energies corresponding to those expected from \ce{^238U} \alphan\ reactions.  The AmBe MC samples systematically produce more background in the neutrino ROI than the PMT neutron MC samples, by \SI{20}{\percent}, arising from differences in emitted neutron energy spectra and positions at which the neutrons were generated. 
We apply a correction to account for this difference and assign a systematic uncertainty equal to the size of the correction. 
As a cross-check to using the AmBe data to evaluate the background, we compare it to the direct MC prediction, and to the result of a multi-component fit to the individual gamma capture lines using the Bayesian Analysis Toolkit~\cite{BAT,Rethmeier:2021oez}.
These methods predict respectively \radiogenicBGMC\ and \radiogenicBGBAT\ events in the ROI, in approximate agreement with the radiogenic \ngamma\ background given in Table~\ref{tab:backgrounds_table}. 

The residual cosmic muon flux at SNOLAB is \muonFluxSNO~\cite{PhysRevD.80.012001}. These muons produce electromagnetic and hadronic showers in the surrounding rock and detector materials. Secondary particles originating from these showers, such as energetic neutrons, \grs, pions, kaons, electrons, positrons can subsequently produce a neutrino-like signal in the LAr even when the primary muons did not pass through the veto detector. Furthermore, muons can activate isotopes with half lives up to \SI{\sim 10}{\second} which can decay outside of the veto time window and lead to backgrounds. We simulated cosmogenic muon-induced backgrounds using RAT~\cite{RAT} and FLUKA~\cite{ballarini2024fluka,Ferrari:898301}. 
The number of Cherenkov photons generated in the \MV\ is recorded for each event, and events are vetoed with a probability calculated based on the number of photons that were created. The probability of the muon veto triggering is obtained with an optical simulation using RAT. Systematic uncertainties on the veto efficiency as a function of the number of Cherenkov photons are evaluated by varying key optical parameters such as the quantum efficiency of the R1408 MV PMTs, the water absorption length, and the tank liner reflectivity by \SI{\pm30}{\percent} relative to their nominal values. Before cuts are applied, in the \SIrange{10}{20}{\MeV} reconstructed energy range we find 24 events tagged by the MV, consistent with the prediction of 20 events.
None of these 24 events pass the \texttt{Low-level cuts}.
The cosmogenic background estimate is given in Table~\ref{tab:backgrounds_table}. 
Because the samples generated with FLUKA did not include a full detailed detector response simulation, this estimate does not include the complete event selection presented in Table~\ref{tab:cut_acceptance}, and the resulting cosmogenic background estimate is conservative.

Random coincidence of events with energies below the ROI may sum to appear within the ROI. These pile-up events include correlated and uncorrelated \alpds\ and \grs. 
Accounting for the \SI{10}{\micro\second} event window, rates for the four dominant sources of pile-up backgrounds are negligible in the ROI, as shown in Table~\ref{tab:backgrounds_table}.

DEAP-3600 is sensitive to high-energy solar neutrinos from the \beight\ and \hep\ reactions. The SNO experiment reported a total flux of active neutrino flavors from \beight\ decays in the Sun to be  \neutrinoFluxSNOthreePhase\, based on a combined analysis of all three phases of solar neutrino data~\cite{PhysRevC.88.025501}. 
Since the \hep\ component has not been measured, we rely on the GS98 solar model~\cite{RevModPhys.92.045006}, which predicts a total flux of \neutrinoFluxhep. Since the charged-current interaction involves only electron neutrinos, we account for the survival oscillation probability as discussed in Ref.~\cite{DENTON2025139560}, using oscillation parameters from NuFIT 6.1~\cite{Esteban2024}. Applying these neutrino fluxes and the cross sections from Bhattacharya \textit{et al.}~\cite{PhysRevC.80.055501}, neutrino interactions were simulated in MARLEY (version 1.2.0)~\cite{GARDINER2021108123}, predicting a total of \nuSigExpectedboroneight\ events from \beight\ neutrinos and \nuSigExpectedhep\ events from \hep\ neutrinos within the ROI. Notably, alternative models utilizing $\beta^+$ decay data or Shell Model calculations predict significantly lower rates. The \grs\ passing through the 1.64~MeV excited state, as generated by MARLEY, are delayed in the simulation by sampling from a 480~ns mean-lifetime exponential distribution. 

\begin{figure}[ht]
   \centering
   \includegraphics[width=0.5\textwidth]{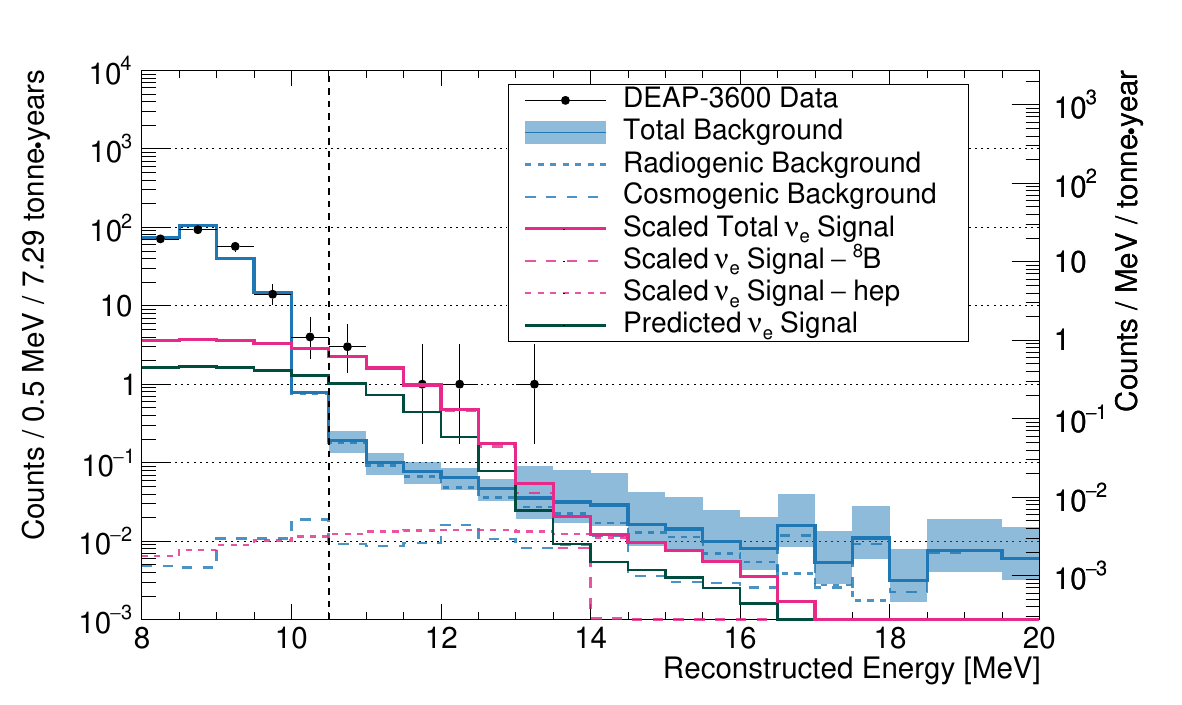}
  \caption{Reconstructed energy distribution for the DEAP-3600 physics dataset (black) with \nuObsExtended~events in the ROI, expected combined background (blue), predicted neutrino signal based on Ref.~\cite{PhysRevC.80.055501} (green), neutrino signal scaled to the measured energy-averaged cross section (magenta). 
  The event in the 13.0--13.5 MeV bin is a double-peak neutrino-like event reconstructing at 13.1 MeV, accepted in our ROI due to the non-linearity of the detector response. No events passing all cuts listed in Table \ref{tab:cut_acceptance} are observed above this event.
  Error bars shown on the data points are the 68\% confidence intervals around the bin contents using the Pearson $\chi^2$ distribution.}
   \label{fig:energy_distribution}
\end{figure}
Figure~\ref{fig:energy_distribution} shows the observed data after all cuts, overlaid with the expected neutrino signal, and the combined background distributions comprised of the sum of the radiogenic and cosmogenic background spectra. After applying all cuts, we observe 6 neutrino-like events in the ROI.
We observe 5 single-peak events between \nuROI, and one double-peak event at 13.1 MeV, within the expanded search region for double-peak events. This event has two distinct peaks separated by \aboveOneDelayedT, and we estimate the energy of the second peak to be 1--2 MeV. While a complete reconstruction of double-peak events will be the subject of future work, this event is consistent with the expected signal from a \beight\ solar neutrino passing through the metastable state of \kforty. We therefore report 6 neutrino-like signals above 10.5~MeV, one of which passes through the metastable state. This observation implies a 68\,\% confidence interval of \SIrange{3}{45}{\%} for the branching ratio of the metastable state, consistent with the theoretical prediction of \branchingMetastable~\cite{GARDINER2021108123}. 
We reject the background-only hypothesis of \totBackground\ events with \nuSigExtended\ significance. 
The data are consistent with the signal-plus-background model within \nuSigBGExtented. 
We calculate an energy-averaged cross section $\mathrm \langle\sigma_{CC}^{ROI}\rangle$ of \effectiveXCsixNobsROI\ in the ROI, a factor of \effectiveXCscale\ higher than that predicted in Ref.~\cite{PhysRevC.80.055501}. 

\begin{figure}[htb]
   \centering
    \includegraphics[width=0.49\textwidth]{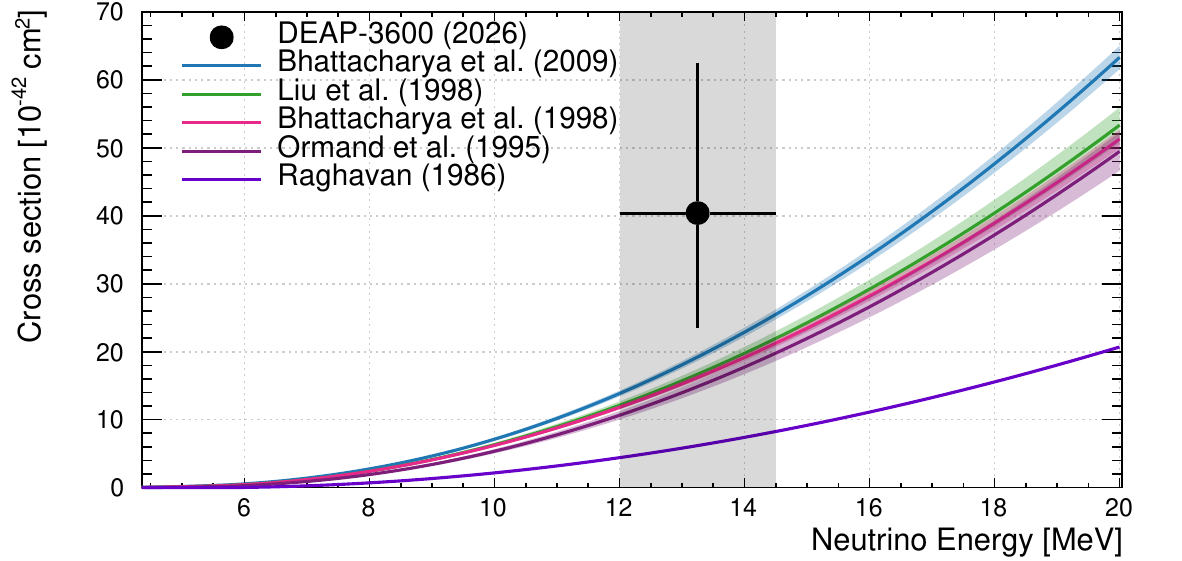}
   \caption{Energy-averaged cross section result, compared with the energy-dependent cross section models Bhattacharya \textit{et al.}~\cite{PhysRevC.58.3677}\cite{PhysRevC.80.055501}, Liu \textit{et al.}~\cite{Liu1998betaDO}, Ormand \textit{et al.}~\cite{ORMAND1995343}, and Raghavan~\cite{PhysRevD.34.2088}.}
   \label{fig:totalXS_results}
\end{figure}

\begin{figure}[htb]
   \centering
    \includegraphics[width=0.42\textwidth]{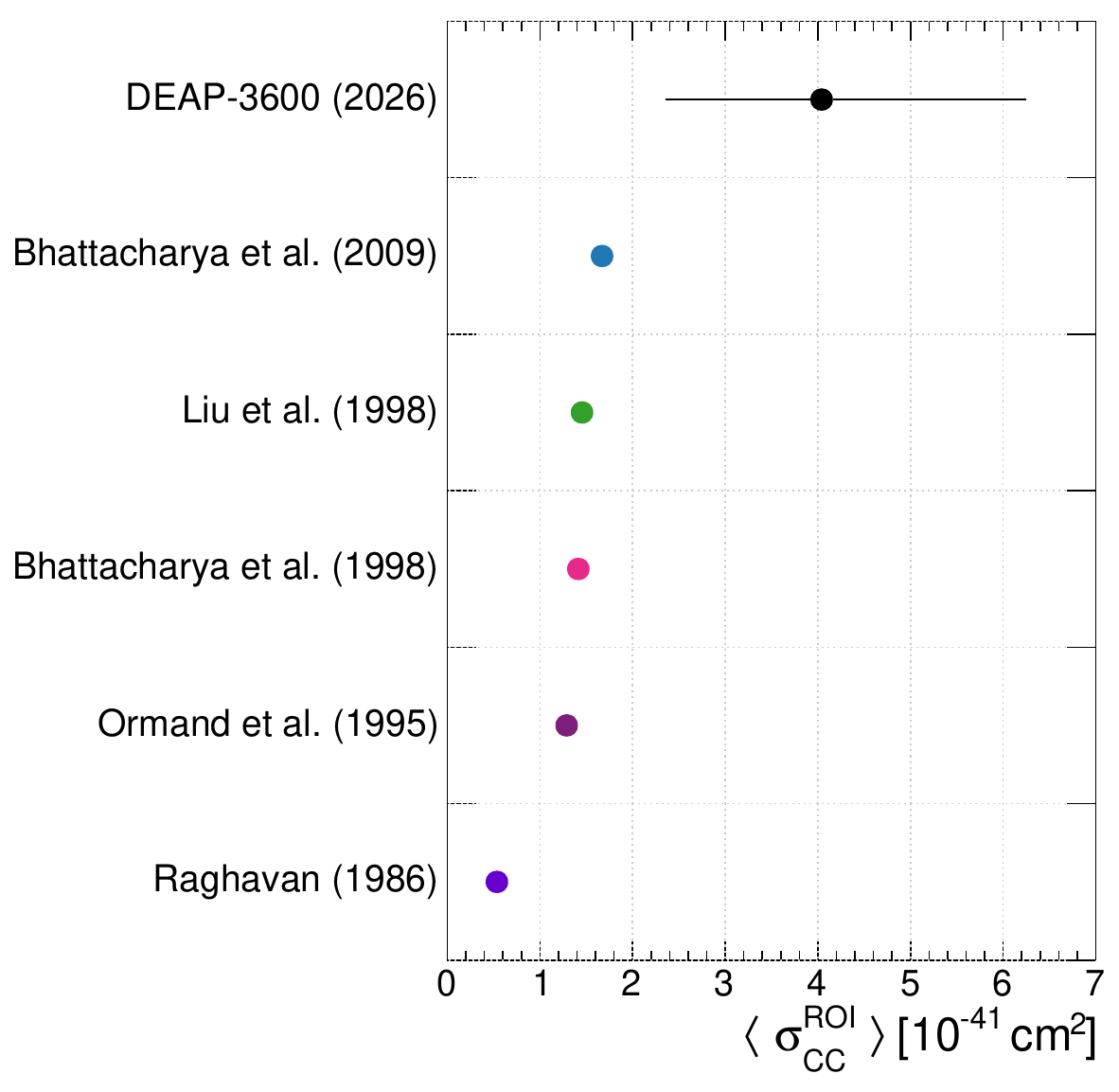}
   \caption{Energy-averaged cross section measurement result in the ROI compared with predictions by Bhattacharya \textit{et al.}~\cite{PhysRevC.58.3677}\cite{PhysRevC.80.055501}, Liu \textit{et al.}~\cite{Liu1998betaDO}, Ormand \textit{et al.}~\cite{ORMAND1995343}, and Raghavan~\cite{PhysRevD.34.2088}.}
   \label{fig:avgXS_results}
\end{figure}

This result is shown in Figure \ref{fig:totalXS_results} with various model predictions for the interaction cross section.
Our energy-averaged cross section measurement is also shown in comparison with the values predicted by other theoretical models in Figure~\ref{fig:avgXS_results}.


\begin{table}[htb]
    \centering
    \caption{The dominant systematic uncertainties considered in this work and their impact on the background and signal estimates in the ROI, further described in the text.
    The energy resolution quoted here is at 10.5~MeV. \texttt{$\langle\sigma_\text{model}\rangle$} is the energy-averaged cross section in the ROI based on Ref.~\cite{PhysRevC.80.055501}.}
    \begin{tabular}{cccc}
    \hline\hline
        {\bf Systematic Unc.} & {\bf Value} & {\bf Bkg} & {\bf CC}~$\mathbf{\nu_{e}}$ \\
        \hline
\texttt{Energy Scale Bias}& 0~$\pm$~1.5\% & 10.1\% & 11.7\%\\
\texttt{Energy Resolution} & 96~keV~$\pm$ 17~keV & 0.5\% & 0.1\%\\   
\texttt{Int./Ext.$\gamma$\,LY\,Bias} & 0~$\pm$~1.3\% & 0.2\% & 12.6\% \\  
\texttt{AmBe Bkg Scaling} & 0.8~$\pm$~0.2& 18.4\% & --\\
\texttt{AmBe Pile-up Est.} & 0.28 $\pm$ 0.09 events& 21.9\% & --\\
\texttt{\MV\ Cut Efficiency}& 99.6\%$^{+ 0.4 \%}_{-1.0\%}$ & ${}^{+14\%}_{-4.4\%}$ & --\\
\texttt{\beight\ solar $\mathrm \nu_e$ flux}& \neutrinoFluxSNOthreePhaseSym & -- & $3.9\%$\\
\texttt{\hep\,solar $\mathrm \nu_e$ flux}& \neutrinoFluxhep & -- & $0.4\%$\\
\texttt{Expected $\langle\sigma_\text{model}\rangle$}  & \effectiveXCBhatt & -- & $3.0\%$\\
    \hline
    \texttt{Total} &  --  & ${}^{+ 33 \%}_{-31\%}$   & 18\% \\
    \hline\hline
    \end{tabular}
    \label{tab:uncertainties}
\end{table}

 The dominant systematic uncertainties in this work are detailed in Table \ref{tab:uncertainties}.
 The uncertainties on \texttt{Energy Scale} and \texttt{Energy Resolution} quantify contributions from uncertainties on the parameters in Eqn.~\ref{eq:erespModel}: 
 these uncertainties in the response function calibrated on single-peak (n,$\gamma$) signals using data taken with AmBe source include statistical fit uncertainties, temporal variations in the detector response between calibration runs, and spatial variations due to source deployment position.
 The systematic difference in light yield between \gr\ events which originate within the LAr to those used for calibration that came from outside the detector is taken as an additional uncertainty (\texttt{Int./Ext.$\gamma$ LY Bias}). 
 The over-prediction of ROI events in the AmBe simulated datasets as compared to the PMT simulated datasets, discussed in regard to radiogenic backgrounds, is also taken as a systematic uncertainty (\texttt{AmBe Bkg Scaling}). 
 The number of pile-up events in the AmBe calibration data is calculated and compared to the direct tagging of double events in the data, with the difference between the quantities taken as a systematic uncertainty (\texttt{AmBe Pile-up Est.}). 
 We also consider the impact of the muon veto (\texttt{\MV\ Cut Efficiency}) on background and signal.

Future work by \DEAP\ will lower the energy threshold by exploiting the delayed metastable transition \gr, and will increase statistics by including upcoming new data runs.

In summary, we report experimental evidence for charged-current interactions (neutrino absorption) of \beight\ solar neutrinos on \arforty.
This measurement establishes LAr as a useful medium by which next-generation multi-tonne scale argon detectors may further study neutrino physics, including improved measurements of the \beight\ solar neutrino spectral shape, determination of neutrino energy spectra produced in supernovae, and direct measurements of \hep\ solar neutrinos. 
Larger detectors with high exposure such as \DSk\ and ARGO are expected to have access to higher precision in the cross section measurements of CC~$\nu_{e}$ and may provide insight on nuclear structure effects, including \kforty~excitation levels and associated transition strengths.

\paragraph{Acknowledgments --}
\begin{acknowledgments}
We thank the Natural Sciences and Engineering Research Council of Canada (NSERC),
the Canada Foundation for Innovation (CFI),
the Ontario Ministry of Research and Innovation (MRI), 
and Alberta Advanced Education and Technology (ASRIP),
the University of Alberta,
Carleton University, 
Queen's University,
the Canada First Research Excellence Fund through the Arthur B.~McDonald Canadian Astroparticle Physics Research Institute,
SECIHTI Project No. CBF-2025-I-1589,
DGAPA UNAM Grant No. PAPIIT IN102326,
the European Research Council Project (ERC StG 279980),
the UK Science and Technology Facilities Council (STFC) (ST/K002570/1 and ST/R002908/1),
the Leverhulme Trust (ECF-20130496),
the Russian Science Foundation (Grant No. 21-72-10065),
the Spanish Ministry of Science and Innovation (PID2022-138357NB-C22),
the Polish National Science Centre (2022/47/B/ST2/02015),
the International Research Agenda Programmes of the Foundation for Polish Science: AstroCeNT (MAB/2018/7), funded from the European Regional Development Fund, and Astrocent (FENG.02.01-IP.05-A015/25), co-financed by the European Union under the European Funds for Smart Economy 2021-2027 (FENG); Teaming for Excellence grant Astrocent Plus (101137080) funded by the European Union with complementary national funding from the Polish Ministry of Science and Higher Education (MNiSW/2025/DIR/811).
Studentship support from
the Rutherford Appleton Laboratory Particle Physics Division,
STFC and SEPNet PhD is acknowledged.
We thank SNOLAB and its staff for support through underground space, logistical, and technical services.
SNOLAB operations are supported by the CFI
and Province of Ontario MRI,
with underground access provided by Vale at the Creighton mine site.
We thank Vale for their continuing support, including the work of shipping the acrylic vessel underground.
We gratefully acknowledge the support of the Digital Research Alliance of Canada,
Calcul Qu\'ebec,
the Centre for Advanced Computing at Queen's University,
the Computational Centre for Particle and Astrophysics (C2PAP) at the Leibniz Supercomputer Centre (LRZ)
and SNOLAB
for providing the computing resources required to undertake this work.
\end{acknowledgments}

\bibliography{References}
\bibliographystyle{apsrev4-1}

\end{document}